\documentclass[twocolumn,superscriptaddress]{revtex4-1}%
\usepackage{amsmath}%
\usepackage{amsfonts}%
\usepackage{amssymb}%

\usepackage{graphicx}

\usepackage{fullpage}

\usepackage{dsfont}
\usepackage{hyperref}
\usepackage{bm}
\usepackage{multirow}
\usepackage{color}
\usepackage{float}

\begin{document}

\title{Strain-dependent exciton diffusion in transition metal dichalcogenides}

 \author{Roberto Rosati}
  \email{roberto.rosati@chalmers.se}
 
 \author{Samuel Brem}
\author{Ra\"ul Perea-Caus\'in}
 
\affiliation{Chalmers University of Technology, Department of Physics,
 412 96 Gothenburg, Sweden}

 \author{Robert Schmidt}
 \author{Iris Niehues}
 \author{Steffen Michaelis de Vasconcellos}
 \author{Rudolf Bratschitsch}
 
 \affiliation{Institute of Physics and Center for Nanotechnology, University of M\"unster, 48149 M\"unster, Germany}
 
 \author{Ermin Malic}
 
 \affiliation{Chalmers University of Technology, Department of Physics,
 412 96 Gothenburg, Sweden}

 \begin{abstract}
 Monolayers of transition metal dichalcogenides (TMDs) have a remarkable excitonic landscape with deeply bound bright and dark exciton states. 
Their  properties are strongly affected by lattice distortions that can be created in a 
controlled way via strain. 
Here, we perform a joint theory-experiment study investigating exciton diffusion in strained tungsten disulfide (WS$_2$) monolayers. 
We reveal a non-trivial and non-monotonic influence of strain. Lattice deformations give rise to different energy shifts  for bright and dark excitons changing the excitonic landscape, 
the efficiency of intervalley scattering channels, and the weight of single exciton species to the overall exciton diffusion. 
We predict a minimal diffusion coefficient in unstrained WS$_2$ followed by a steep speed-up by a factor of 3 for tensile biaxial strain at about 0.6\% strain -  
in excellent agreement with our experiments.
   The obtained microscopic insights on the impact of strain on exciton diffusion are applicable to a broad class of multi-valley 2D materials.
   \end{abstract}
\maketitle

\begin{figure}[t!]
\centering
\includegraphics[width=\linewidth]{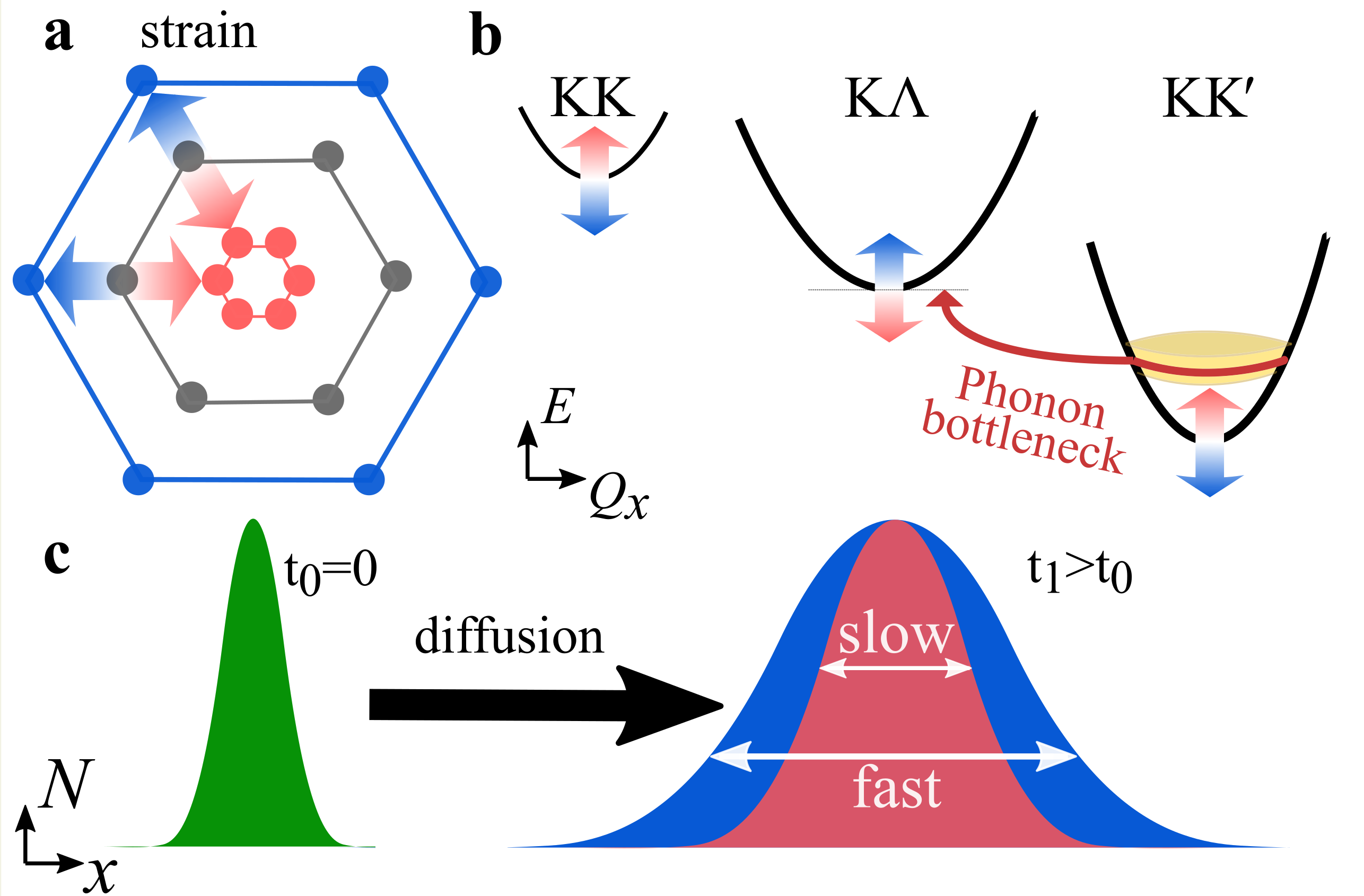}
\setlength{\belowcaptionskip}{-10pt}
\setlength{\abovecaptionskip}{-10pt}
\caption{\textbf{Strain-dependent exciton diffusion.} Biaxial
strain implies (a) lattice deformations, which modify
(b) the excitonic landscape by different spectral shifts of the bright KK and momentum-dark  K$\Lambda$ and KK$^\prime$ excitons. This can introduce strain-dependent phonon bottlenecks for intervalley scattering. (c) As a result, an initial
spatially-localized exciton occupation diffuses slower or faster depending on the applied strain. 
\label{fig:Fig1}}
\end{figure}

Transition metal dichalcogenides (TMDs) and related van der Waals heterostructures have attracted much attention in current research due to their 
remarkable excitonic landscape including bright, spin- and momentum-dark, and
spatially separated exciton states \cite{Wang18,Mueller18,Malic18,Mak10,
Chernikov14, Kunstmann18,Merkl19, Brem20b}
Optical properties of TMDs are strongly sensitive to lattice distortions, as shown for spatially-homogenous \cite{Schmidt16,Conley13,He13,Plechinger15,
Frisenda17,Niehues18,Peng19,Mennel20} and spatially-inhomogeneous strain \cite{Tonndorf15,Kern16,Palacios17,Branny17,Rosenberger19,Tyurnina19,Carmesin19,Harats20}.
Strain-induced lattice deformations [see Fig. \ref{fig:Fig1}a] induce changes in the energy and in the effective mass of electronic valleys 
 \cite{Feng12,He13,Steinhoff14,Khatibi18,Feierabend17b} resulting in a qualitative change of the excitonic landscape and efficiency of exciton-phonon scattering channels
 (Fig. \ref{fig:Fig1}b)\cite{Niehues18,Khatibi18,Feierabend17b}.

Strain  is expected to have an important impact also on transport in TMDs, e.g. spatially inhomogenous strain acts as a driving force for 
exciton/carrier funneling  \cite{Cordovilla18,Tyurnina19,Harats20},
similarly to bias fields for charged particles \cite{Wang12,Mak16} or gauge potentials for interlayer excitons in van der Waals heterostructures \cite{Unuchek18}. 
Even in the absence of additional driving forces, an initial spatially-localized excitonic 
occupation  spreads in space due to the
occupation gradient resulting eventually in a conventional exciton diffusion \cite{Rosati20}, as sketched in Fig. \ref{fig:Fig1}c.
While in the presence of one single valley the diffusion is expected to be only quantitatively altered by strain through e.g. changes in the effective masses, the multi-valley exciton landscape in TMDs promises interesting strain effects.  The diffusion is expected to be dominated by 
the most populated, energetically lowest exciton valley \cite{Rosati20,Zipfel20,Perea19}, which might vary as a function of strain due to the strongly valley-dependent energy shifts 
[Fig. \ref{fig:Fig1}b].  
Since different valleys have different valley-intrinsic diffusion coefficients \cite{Rosati20}, this leads to strain-induced changes of the overall diffusion, see Fig. \ref{fig:Fig1}c.
Furthermore, strain-induced energy shifts also change the possibility for phonon-induced intervalley scattering channels and may even result in phonon bottlenecks at specific strain values  [Fig. \ref{fig:Fig1}b]. This has also a direct impact on the efficiency of exciton diffusion.

Based on a fully quantum-mechanical approach and supported by experimental measurements, our work provides microscopic insights into the interplay of  exciton diffusion and strain in the WS$_2$ monolayer as an exemplary TMD material. 
We microscopically address the evolution of optically excited, spatially localized excitons resolved in time, momentum, and space. We take into account bright and momentum-dark excitonic states obtained by solving the Wannier equation 
under strain \cite{Feierabend17b,Khatibi18}.
We predict non-trivial dependence of the diffusion on strain, showing a non-monotonic behaviour, where the overall diffusion
is either dominated by specific dark excitons or determined by intervalley scattering. This leads to a steep speed-up of the diffusion upon small tensile strain values - in excellent agreement with our spatiotemporal photoluminescence experiments.

\section{Results}
\textbf{Theoretical approach:}
Starting with the unstrained single-particle dispersion \cite{Kormanyos15}, we implement the strain-induced variations of effective masses and band extrema $E_v$  \cite{Khatibi18}. For each strain value we then solve the Wannier
equation  \cite{Haug09,Selig16,Selig18,Brem18} 
 with a non-local Coulomb screening \cite{Brem19b} to obtain a set of excitonic states 
$\vert\alpha\rangle\equiv\vert\mathbf{Q},v\rangle$ labelled by the excitonic valley $v$ and the center-of-mass momentum $\mathbf{Q}$.
These states have the energy $\varepsilon_\alpha=E_v+\hbar^2|\mathbf{Q}|^2/(2M_v)$, 
which depends on strain largely via changes in $E_v$ \cite{Khatibi18} [see also Fig. \ref{fig:Fig1}b], but also via changes in the total mass $M_v$. 
Due to considerable energetic separations to higher excitonic states, 
we restrict our investigations to the ground $1s$ exciton, however taking into account all relevant electronic valleys and the resulting bright KK as well as momentum-dark KK$^\prime$, K$\Lambda$, K$\Lambda^\prime$, $\Gamma$K and $\Gamma$K$^\prime$ excitons \cite{Malic18}. 
Here, the first and the second letter denote the location of the Coulomb-bound hole and electron, respectively. 

Now, we introduce the excitonic intravalley Wigner function 
$N^v_{\mathbf{Q}}(\mathbf{r},t)$, which summed over $\mathbf{Q}$ provides the 
intravalley spatial density 
$N_v(\mathbf{r},t)\equiv \frac{1}{V}\sum_{\mathbf{Q}} N^v_{\mathbf{Q}}(\mathbf{r},t)$ \cite{Rosati20}. 
At the spatial and temporal scales considered here, $N^v_{\mathbf{Q}}(\mathbf{r},t)$ can be directly
interpreted as probablity of finding excitons with momentum $\mathbf{Q}$ in position $\mathbf{r}$ and valley $v$.
An equation of motion
for the spatiotemporal dynamics of excitons can be introduced by exploiting the Heisenberg equation \cite{Rosati20}. 
The derived  equation can 
then be transformed into  Wigner representation \cite{Hess96,Jago19} and reads in the low excitation regime
\begin{align}
\label{SBE}
\nonumber
\dot{N}^v_\mathbf{Q}(\mathbf{r},t)&= \left(\frac{\hbar \mathbf{Q}}{M_v}\cdot \nabla - \gamma \delta_{\mathbf{Q},0}\delta_{v,KK} \right)N^v_\mathbf{Q}(\mathbf{r},t)\\ 
&+\Gamma^{v;KK}_{\mathbf{Q};0} |p_{0}(\mathbf{r},t)|^2\!\!+\!\!\left.\dot{N}^v_\mathbf{Q}(\mathbf{r},t)\right|_{sc}\hspace{-5pt}.
\end{align}
The first term indicates the free evolution of excitons, while the second term takes into account the losses due 
to the radiative recombination  $\gamma$  within the light cone ($\delta_{\mathbf{Q},0}\delta_{v,KK}$) \cite{Selig16,Selig18, Brem18, Brem19}. 

The
first contribution in the second line of Eq. (\ref{SBE}) describes the formation of incoherent excitons due to phonon-driven transfer from 
the excitonic polarization $p_{\mathbf{Q}\approx 0}(\mathbf{r}, t)$. 
The process is driven by exciton-phonon scattering  rates $\Gamma^{vv^\prime}_{\mathbf{Q}\mathbf{Q}^\prime}$ describing scattering from the state $\vert \mathbf{Q}^\prime v^\prime\rangle$ to $\vert \mathbf{Q} v\rangle$  via interaction with phonons \cite{Selig18, Brem18}.
The last term in Eq. (\ref{SBE}) describes the scattering contribution $\left.\dot{N}^v_{\mathbf{Q}}(\mathbf{r},t)\right|_{sc}=\Gamma^{\text{in},v}_{\mathbf{Q}}(\mathbf{r},t)\!-\!\Gamma^{\text{out},v}_{\mathbf{Q}}N_{\mathbf{Q}}(\mathbf{r},t)$, which is dominated by exciton-phonon scattering in the considered low-excitation regime. It  gives rise to a redistribution
 of the Wigner function in momentum toward a local equilibrium  distribution, cf. the supplementary material for more details. 
 Importantly, exciton-phonon scattering depends crucially on strain, mostly via the energies of the involved initial and final exciton states, while the variations
of phonon energies or electron-phonon scattering are typically less relevant and are thus not considered here \cite{Khatibi18,Dadgar18}. 
In particular, strain-induced energy shifts can lead to drastic changes in intervalley scattering via opening or closing of 
 scattering channels [see phonon bottleneck in Fig. \ref{fig:Fig1}b]. In contrast, intravalley scattering  is only slightly influenced by strain via changes 
in the effective mass $M_v$ \cite{Khatibi18,Aslan18,Aslan20}. 

 In the steady-state regime 
the interplay between scattering-free propagation and scattering mechanisms 
implies that the spatial distribution
evolves at first approximation according to the  Fick's law
$\dot{N}_v(\mathbf{r},t)=D_v \Delta_{\mathbf{r}} N_v(\mathbf{r},t)$, where decaying mechanisms have been omitted \cite{Kato16,Cadiz17}. 
Here, $D_v=1/2\langle \tau^v_{\mathbf{Q}}\hbar^2 Q^2/M_v^2 \rangle\vert^v_{\mathbf{Q}}$ is the diffusion coefficient 
 with $1/\tau^v_{\mathbf{Q}}=\sum_{\mathbf{Q}^\prime,v^\prime} \Gamma^{v^\prime v}_{\mathbf{Q}^\prime\mathbf{Q}}$ 
denoting the momentum-dependent relaxation times. We have introduced
the expectation value 
$\langle f_{\mathbf{Q}}\rangle\vert^v_\mathbf{Q}=\sum_{\mathbf{Q}}f_{\mathbf{Q}}\textrm{Exp}
\big(-\frac{\varepsilon_{\mathbf{Q}v}}{k_BT}\big)/\sum_\mathbf{Q}\textrm{Exp}\big(-\frac{\varepsilon_{\mathbf{Q}v}}{k_BT}\big)$. Under the assumption of constant relaxation times 
$\tau^{v}_{\mathbf{Q}}\approx \tau_v$ the well-known steady-state  relation $D_v=\tau_v k_B T/M_v$ can be recovered. 
The overall diffusion of the total excitonic spatial density $N=\sum_v N_v$ can be affected by strain in two major ways: (i) Strain changes the relative occupation weight $N_v/N$ of each valley. The overall diffusion will be dominated by the most populated valley and its own valley-intrinsic
diffusion \cite{Rosati20}. (ii) Strain changes the efficiency of the scattering channels through shifting the energies of initial and final scattering states, which crucially
determine the velocity of  exciton diffusion.

\begin{figure}[t]
\centering
\includegraphics[width=\linewidth]{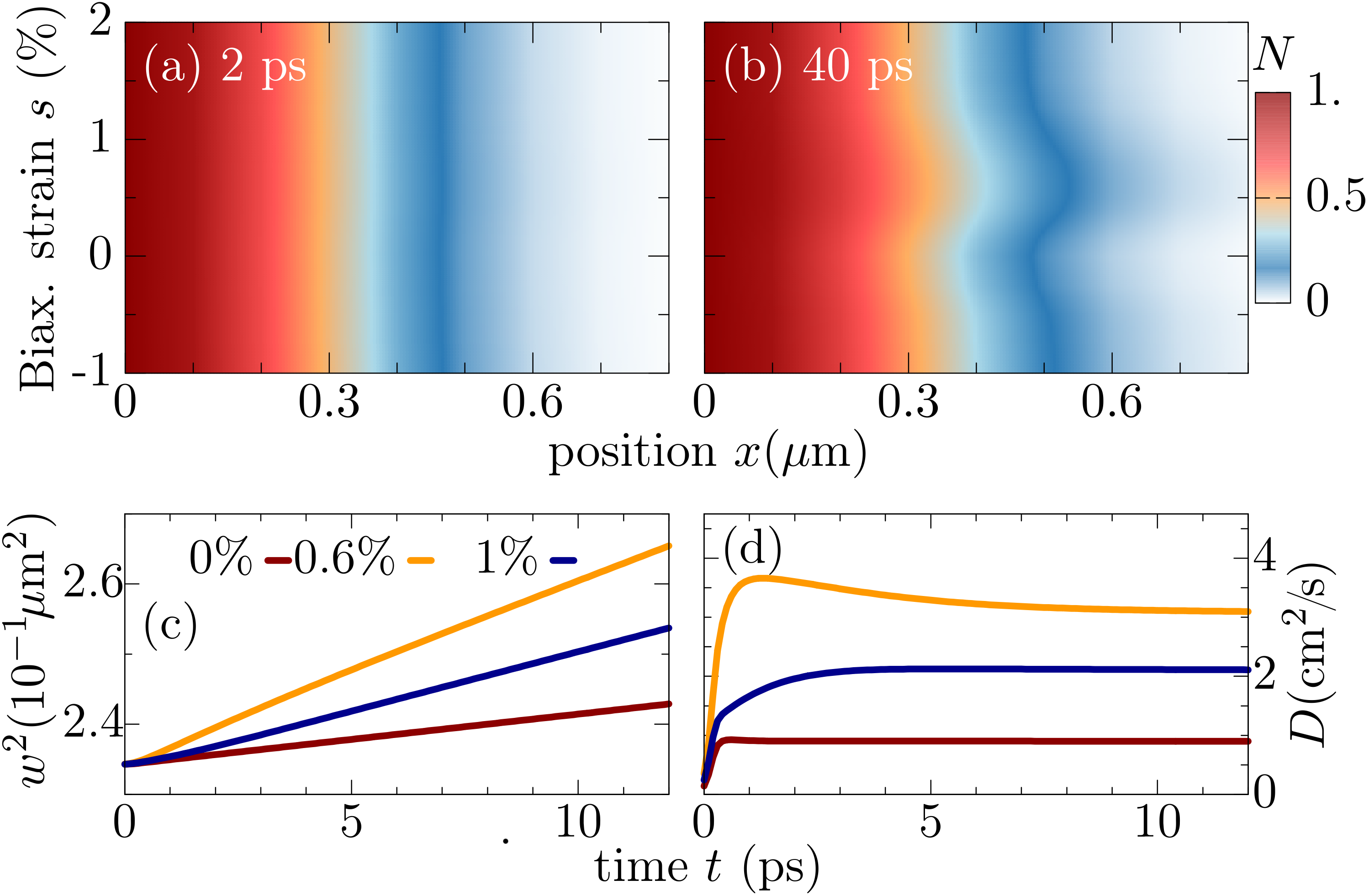}
\caption{\textbf{Spatiotemporal evolution under strain.} 
Strain-dependent spatial evolution of the exciton density $N$
in a WS$_2$ monolayer on SiO$_2$ after a pulsed spatially-localized
optical excitation. The spatial evolution is shown along the $x$
axis  at the fixed time of (a) 2 ps soon after the optical excitation and after (b) 40 ps. For given strain values, we show
the evolution of the corresponding (c) squared spatial width $w^2$ and the (d) associated effective diffusion coefficient $D$. 
\label{fig:Fig2}}
\end{figure}

\textbf{Strain-dependent exciton diffusion:} Exploiting Eq. (\ref{SBE}), 
we have a microscopic access to the spatially and temporally resolved dynamics of excitons in strained TMDs. In the main manuscript, we focus on the exemplary case of WS$_2$ monolayers on a SiO$_2$ substrate. In supplementary material, we also discus the exciton diffusion in strained WSe$_2$, MoSe$_2$, and  MoS$_2$ monolayers.
Figures \ref{fig:Fig2}(a)-(b) illustrate the strain dependent spatial evolution of the exciton density $N(\mathbf{r},t)$ soon after the optical excitation [$t$=2 ps in (a)] and
at a later time [$t$=40 ps in (b)]. We consider a pulsed optical excitation around $t=0$ resonant 
to the bright exciton X$_0$, and with a Gaussian spatial confinement  corresponding to a full-widths-half-maximum of 800 nm in amplitude and a  temporal duration of 200 fs. This generates an initial strain-independent excitonic distribution, cf. Fig. \ref{fig:Fig2}a. After few tens of picoseconds, the spatial distribution becomes broader and the width is strongly strain dependent, cf. Figs. \ref{fig:Fig2}(b). We find in particular a faster spatial spreading for given strain values $s$, e.g. $s\approx 0.6\%$. The differences in exciton diffusion
are not monotonic in strain: Increasing the strain from negative (compressive) to positive (tensile) values, the diffusion initially becomes slower from $-1\%$ to $0\%$ strain, then it speeds up steeply as the strain increases to $0.6\%$ and finally it slows down again for larger strain values.

The spatial broadening of the exciton density $N$ can be quantified  introducing a width $w$ whose squared modulus is proportional to the variance $w^2= \int \mathbf{r}^2N(\mathbf{r},t) d\mathbf{r}/N$. According to Fick's law, confined spatial distributions behave as $N(\mathbf{r},t)\propto \exp\left[-r^2/w^2(t)\right]$ with $w^2(t)=w_0^2+4 D t$ \cite{He15,Kulig18}, where $w_0$ is the initial width.  
It follows that also in the microscopic case when evaluating Eq. (\ref{SBE}) one can define an \textit{effective diffusion coefficient} (also called \textit{diffusivity})  $D=\frac{1}{4}\partial_t w^2$, i.e. as  slope of the temporal evolution of the squared width $w^2$.  
In Figs. \ref{fig:Fig2}(c) and (d) we plot the temporal evolution of squared width $w^2$ and the associated
effective diffusion coefficient for three different values of strain. 
The squared width $w^2$ shows quickly a linear evolution for all strain values, although the slope varies  crucially with strain indicating a strong strain-dependence of the exciton diffusion. We find that  after an initial steep increase, the diffusion coefficient $D(t)$ reaches a stationary value after a few ps. 
This corresponds to the transition from a ballistic regime, i.e. scattering-free evolution with a quadratic $w^2$ and linear $D$,  to the
conventional diffusive regime with a linear $w^2$  and stationary $D(t)\equiv D$ \cite{Rosati20}.  
Both strain-induced quicker ballistic-to-diffusive transition and smaller $D$ values are  signatures of more efficient scattering channels, as seen e.g. at 0$\%$ strain.
Due to the very short ballistic-to-diffusive transition, it is the stationary
diffusion coefficient $D$ that determines the exciton diffusion in Fig. \ref{fig:Fig2}b.

\begin{figure}[t!]
\centering
\includegraphics[width=\linewidth]{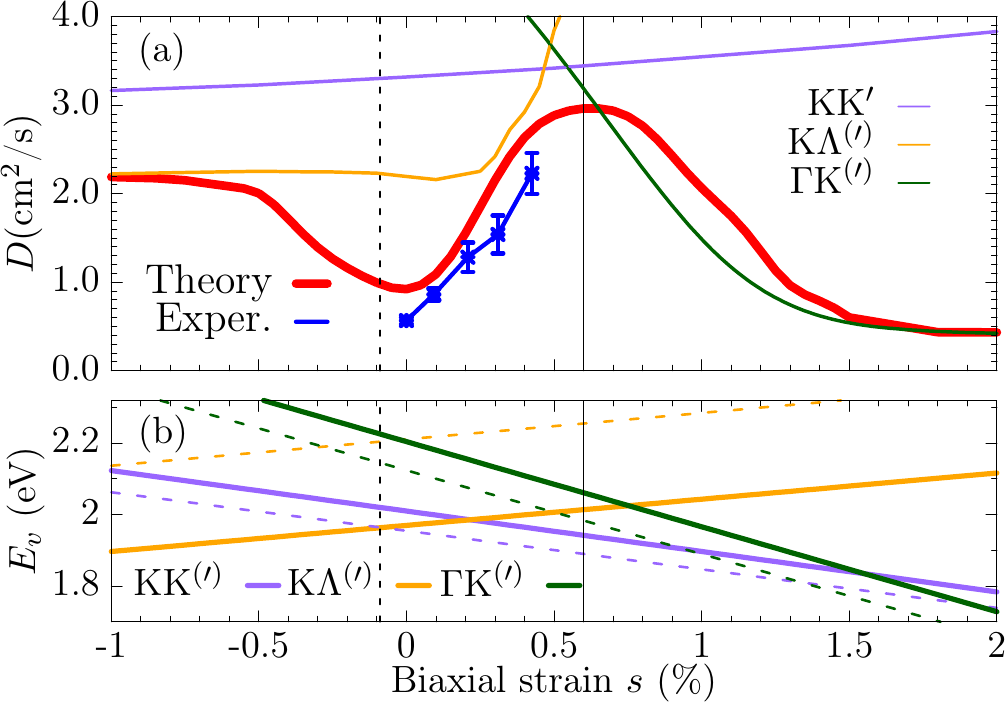}
\caption{\textbf{Strain-dependent stationary diffusion.}
(a) Diffusion coefficient plotted as a function of  biaxial strain in WS$_2$ containing a direct comparison between theoretically predicted (red solid line) and experimentally measured (blue line with errorbars) values. 
Note that the experiment was performed with uniaxial strain that has been projected to equivalent biaxial strain values (cf. supplementary material).
The thin lines show the gedanken experiment where only KK$^\prime$
(blue), K$\Lambda^{(\prime)}$
(orange) and $\Gamma \text{K}^{(\prime)}$
(green) excitons are taken into account (besides the optically excited KK states). (b) Strain-dependent energy shifts of the minima of KK$^{(\prime)}$, K$\Lambda^{(\prime)}$ 
and K$\Gamma^{(\prime)}$ excitons. The dashed lines indicate the case with electrons located in equivalent primed exciton valleys.
The vertical dashed line marks the strain at which  $E_{\text{K}\Lambda}\approx E_{\text{KK}^\prime}$ giving rise to a minimum in the diffusion coefficient, 
while the vertical solid line indicates the maximum of the diffusion. 
\label{fig:Fig3}}
\end{figure}

Now, we investigate the strain-dependence of the stationary diffusion coefficient covering a larger range of compressive to tensile biaxial strain values, cf.  Fig. \ref{fig:Fig3}a. 
We find saturation values for the diffusion coefficient of 2.2 and 0.4 cm$^2$/s for compressive strain around -1\%  and tensile strain above +1.6\%, respectively. 
Furthermore, we predict a strongly non-monotonic strain-dependence including a relative minimum in the unstrained material (0\% strain) and a maxium at about 0.6 $\%$ tensile strain. This can be understood by decomposing the valley-dependent contribution to the overall exciton diffusion. 
Thin lines in Fig. \ref{fig:Fig3}a show the diffusion coefficient obtained considering a reduced excitonic valley landscape. 
In all cases we have taken into account the optically excited KK excitons  
together with only  KK$^\prime$ (purple line) or
  K$\Lambda$ and K$\Lambda^\prime$ (i.e. K$\Lambda^{(\prime)}$, orange line) or $\Gamma$K$^{(\prime)}$ (green line) excitons, respectively. 
These thin lines reflect the scenario of valley-intrinsic diffusion, where the overall diffusion is dominated by the considered specific valley and 
the corresponding valley-specific diffusion determined by its total mass and intravalley scattering \cite{Rosati20}. We find immediately that the saturation value of the diffusion at compressive (tensile) strain is determined by K$\Lambda$ ($\Gamma$K) excitons, while the maximum in diffusion is governed by the contribution of KK$^\prime$ exciton.

To better understand the valley-specific contribution to the overall diffusion, we show the strain dependent  energy of the involved exciton valley minima in Fig. 
 \ref{fig:Fig3}b \cite{Khatibi18}. 
We find that at larger compressive (tensile) strain values,  K$\Lambda$ ($\Gamma$K$^{(\prime)})$  excitons are the energetically lowest and thus most occupied states. 
The same applies to KK$^\prime$ excitons for strain values around 0.5\%. As shown in our previous work \cite{Rosati20}, 
efficient intervalley scattering gives rise to one joint  diffusion coefficient, however the weight of each valley is determined  by its relative occupation. 
The latter depends crucially on the position of the bottom of the valley $E_v$ as well as on degeneracy (3 times larger for K$\Lambda^{(\prime)}$) or total mass (smaller total mass leads
to smaller population): This explains why  K$\Lambda$ or the very massive $\Gamma$K$^{(\prime)}$ excitons still contribute even when 
$E_{\text{KK}^\prime}<E_{\text{K}\Lambda},E_{\Gamma\text{K}^\prime}$ [see e.g. vertical solid line in Fig. \ref{fig:Fig3}a].
At very strong compressive strain it is K$\Lambda$ which dominates the diffusion. 
Since at these strain values, the K$\Lambda$ valley is already by far the energetically lowest state, the diffusion coefficient is not affected anymore by 
further strain-induced energetic changes resulting in a stationary value. The same also applies to the situation at high tensile strain, where the $\Gamma\text{K}^\prime$ 
valley when low enough becomes dominant also due to its large effective mass and high exciton occupation.

The predicted maximum in the exciton diffusion at about 0.6\% is formed when moving from the strain regime, where  KK$^\prime$ excitons are energetically lowest states toward the 
regime governed by $\Gamma$K$^{(\prime)}$ states. The latter exhibit a considerably slower diffusion due to a much larger effective mass. 
The dip toward the unstrained case cannot be explained by just considering valley-intrinsic diffusion and is due to intervalley scattering, which will be discussed separately in the  next section. 
Note that the abrupt increase in the diffusion coefficient in the considered hypothetical two-valley system dominated by $\Gamma$K$^{(\prime)}$ (thin green line) 
and K$\Lambda$ (thin orange line) excitons for strain values around 0.5\% is due to the increasing impact of KK excitons exhibiting  
a much smaller effective mass and thus a much larger diffusion coefficient \cite{Rosati20}. The thin purple line does not show this steep increase, since
KK$^\prime$ and KK excitons have a very similar dependence on strain modifying their separation by only approximately 5 meV per percentage of biaxial strain (cf. solid and dashed purple line in Fig. \ref{fig:Fig3}b).

We compare our microscopic results with measurements of the exciton diffusion in an uniaxially strained WS$_2$ monolayer. 
To this end, the monolayer is placed on a flexible PMMA substrate and a homogeneous, uniaxial, tensile strain from 0~\% to 1~\% is applied via the bending method \cite{Schmidt16, Niehues18}. 
The similar optical properties between PMMA and SiO$_2$  make a comparison of experiment and theory possible.  Since we apply uniaxial strain, the experimentally determined strain values are related to the biaxial ones in the theory by comparing the energy shifts of the KK$^{(\prime)}$. 
The exciton diffusion is measured by space- and time-resolved photoluminescence after applying a spatially localized optical excitation at 2.10 eV  (see supplementary information for details).
We find an excellent agreement between theoretically predicted and experimentally measured diffusion coefficients in the experimentally accessible strain region, cf. Fig. \ref{fig:Fig3}a. 
We see an increase in the diffusion coefficient from approximately 0.6 to 2.2 cm$^2$/s when varying the strain from 0 to 0.5\%. 	
The slope of this increase strongly depends on the
formation of the minimum 
in the diffusion coefficient $D$ 
appearing approximately for the unstrained case.

\textbf{Impact of intervalley scattering:}
While the overall diffusion discussed so far was dominated by one exciton valley with the lowest energy and largest occupation (valley-intrinsic diffusion), there are strain regions where multiple valleys  have similar energies, cf. Fig. \ref{fig:Fig3}b.  This occurs e.g. at small compressive strain values, where K$\Lambda$ (solid orange) and KK$^\prime$ excitons (dashed purple) cross, cf.  the vertical dashed line. 
Here, intervalley scattering turns out to play a crucial role resulting in a minimum in the exciton diffusion, cf. Fig. \ref{fig:Fig3}a.

\begin{figure}[t!]
\centering
\includegraphics[width=\linewidth]{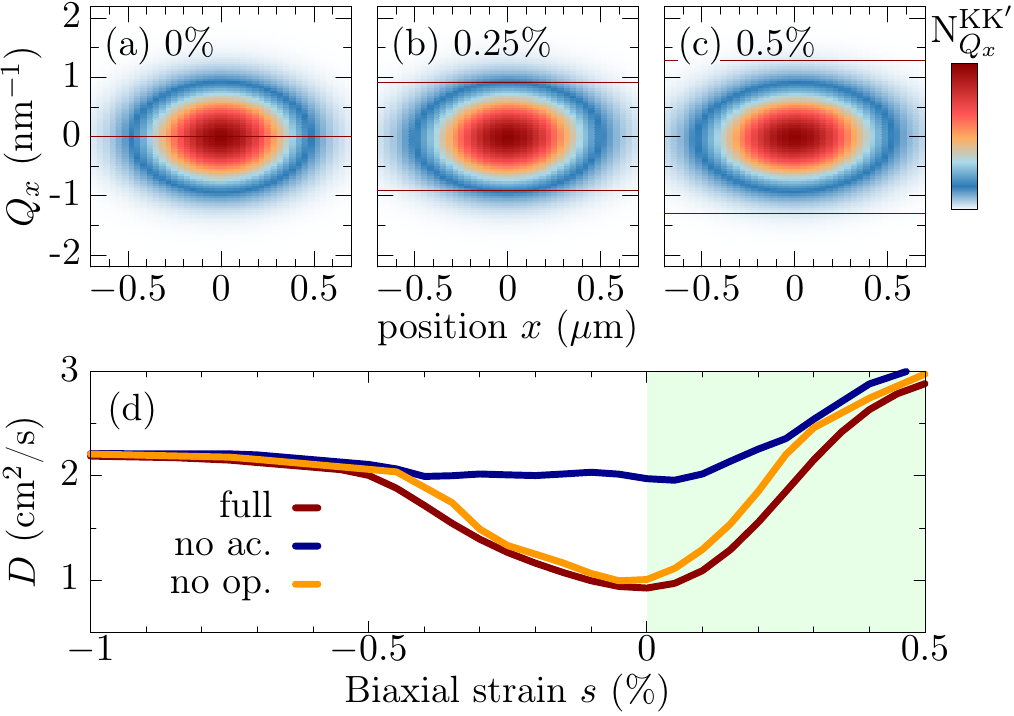}
\caption{\textbf{Impact of intervalley scattering.}
(a)-(c) Wigner function $N^v_{Q_x}(x)$ is shown in dependence on $\mathbf{Q}\equiv(Q_x,0)$ and $\mathbf{r}\equiv(x,0)$ at 40 ps for the  
lowest  $v=$KK$^\prime$ valley and three different values of strain. Thin horizontal lines indicate the minimum wave-vector required for intervalley scattering via TA phonons. 
 (d) Comparison of the full simulation (dark-red) with the cases in 
which only acoustic (dark-blue) or optical (orange) phonons are switched off.
\label{fig:Fig4}}  
\end{figure}

Now, we investigate the origin of the predicted dip in the diffusion coefficient  for the unstrained WS$_2$ monolayer, cf. Fig. \ref{fig:Fig3}a. 
In particular, we study the role of intervalley scattering between KK$^\prime$ and K$\Lambda$ excitons,  whose energies are the lowest and cross in the strain region, where the minimum appears, cf. Fig. \ref{fig:Fig3}b. 
The intervalley scattering between KK$^\prime$ and K$\Lambda$ excitons is driven by absorption or emission of high-momentum 
M phonons. These phonons can be approximated as nearly dispersion-free including
two acoustic modes with the energies of 16.5 and 22.7 meV. The corresponding optical M phonons have larger energies between 40 and 50 meV \cite{Jin14}. 
When the strain-induced separation between valleys becomes larger than these phonon energies, 
excitons from the energetically lowest states cannot scatter any longer out of these states, cf.  Fig. \ref{fig:Fig1}b. 
To  better visualize the condition for this phonon bottleneck, we show the Wigner function for the KK$^\prime$ excitons for three different strain values, where KK$^\prime$ are the energetically lowest states, cf. Figs. \ref{fig:Fig4}(a)-(c).
Here, the thin horizontal lines show the minimum momentum $\vert \bar{Q}_x\vert$ required for the absorption of acoustic M phonons. In the unstrained case [Fig. \ref{fig:Fig4}a], 
the two momentum-dark exciton valleys are aligned [Fig. \ref{fig:Fig3}b], thus
all intervalley channels into K$\Lambda$ states are possible. The situation changes at 0.25 \% strain [Fig. \ref{fig:Fig4}b], where the energetic misalignment is already large enough 
that only a small portion of the most occupied excitons  is able to scatter into K$\Lambda$ states. At 0.5\% strain [Fig. \ref{fig:Fig4}c], there are almost no occupied states fulfilling the condition for intervalley scattering resulting in a phonon bottleneck. 

Now we address the impact of this phonon bottleneck for the strain-dependent evolution of the exciton diffusion. In Fig. \ref{fig:Fig4}d, we compare the full calculation (also shown in Fig. \ref{fig:Fig3}a)
with the case, where the KK$^\prime$-K$\Lambda$ intervalley scattering via acoustic or optical phonons is switched off, respectively. 
We see that optical phonons have in general a minor effect, since their energy is relatively high resulting in a small phonon occupation and thus a negligbly small phonon absorption. We also find that scattering with acoustic phonons does not play a role for strain values $|s|\gtrsim 0.5\%$. The reason is the phonon bottleneck illustrated in Fig. \ref{fig:Fig4}c. However, in the strain region $|s|\lesssim 0.5\%$, 
switching off scattering with acoustic phonons  suppresses the appearance of the dip in the exciton diffusion. This illustrates the crucial role of intervalley scattering with acoustic phonons for the formation of the predicted minimal exciton diffusion in unstrained WS$_2$ monolayers.  

Note that the scattering with M phonons connecting KK$^\prime$ and K$\Lambda$ excitons is particularly strong according to DFT calculations \cite{Jin14}. 
In contrast, the corresponding phonon-induced intervalley scattering connecting $\Gamma$K$^{(\prime)}$ and KK$^{(\prime)}$ excitons is much weaker. 
This together with the much larger effective mass and smaller diffusion coefficient in the $\Gamma\text{K}^{(\prime)}$ valley does not lead to a similar  
dip in the exciton diffusion around 1.5\% strain, where   KK$^{\prime}$- and  $\Gamma\text{K}^{(\prime)}$ states are energetically lowest and cross [Fig. \ref{fig:Fig3}b].
Note that the efficiency of the KK$^\prime$-K$\Lambda$ intervalley scattering together with the small energy of acoustic M phonon leads to a persistence of the dip also at smaller temperatures, as shown in the supplementary material. 

Finally, we briefly discuss the strain-dependence of exciton diffusion in  other TMD materials. While the physical mechanisms are the same, in Mo-based monolayers one needs a larger tensile or compressive strain to reach the multi-valley 
features discussed for WS$_2$.
In particular, the situation in MoS$_2$ is more involved due to the initial excitonic landscape in the unstrained case, where the position of $E_{\Gamma\text{K}^{(\prime)}}$ strongly
depends  on the relative distance $\Delta E_{\text{K}\Gamma}$ of the valence-band maxima located in K and  $\Gamma$. The latter is still being controversially discussed in literature \cite{Kormanyos15,Khatibi18,Uchiyama19,Deilmann19}.
We predict a qualitatively different  strain-dependent exciton diffusion in MoS$_2$ depending on $\Delta E_{\text{K}\Gamma}$, cf. the supplementary material. An experimental study of exciton diffusion in MoS$_2$ could thus provide a better understanding of the relative position of the bright KK and the dark $\Gamma$K$^{(\prime)}$ excitons.

\section{Discussion}
In conclusion, the presented joint theory-experiment study provides new microscopic insights into 
strain-dependent exciton diffusion in TMD monolayers. 
We find that the diffusion becomes faster or slower with strain in a non-trivial and non-monotonic way. This is a result of the interplay between lattice-distortions and the remarkable multi-valley excitonic 
landscape in TMDs.
Strain-induced shifts of exciton energies change the relative energy separations in the excitonic landscape of bright and momentum-dark excitons. This has an immediate impact on which state is dominant and governs the overall diffusion coefficient. In particular, we predict a dip in the diffusion for unstrained WS$_2$ monolayers that we microscopically ascribe to intervalley scattering with acoustic phonons. 
This dip is followed by a large increase of exciton diffusion by a factor of 3 for tensile biaxial strain of up to 0.6\%. The theoretical prediction is  found to be in excellent agreement with spatiotemporal photoluminescence experiments.
Overall, our study provides microscopic insights into the impact of strain on exciton diffusion in technologically promising 2D materials and uncovers  the underlying fundamental intra- and intervalley scattering processes involving bright and momentum-dark excitons.

\textbf{Supporting Information}

Additonal details on microscopic modeling and measurements on  strain-dependent exciton diffusion are included. Furthermore, temperature dependence  as well as exciton diffusion in different strained TMD monolayers are discussed. 

\textbf{Acknowledgements:}
This project has received funding from the Swedish Research
Council (VR, project number 2018-00734) and the European
Union’s Horizon 2020 research and innovation programme
under grant agreement no. 881603 (Graphene Flagship). 
The authors thank Zahra Khatibi and Maja Feierabend
for fruitful discussions.


%

\end{document}


\maketitle



\section{Wannier equation and exciton-phonon scattering}

The energetic position of the excitonic energies $E_{v}$ is given
by $E_{v}=E_{v}^{0}-E_{v}^{\textrm{b}}$, where $E_{v}^{0}$ is the
single-particle separation between top of the valence band in the
K and $\Gamma$ valley and bottom of the conduction band in the K,K$^{\prime}$
and $\Lambda$ valleys for $v=\text{KK}^{(\prime)},\text{K}\Lambda^{(\prime)},\Gamma\text{K}^{(\prime)}$.
The exciton binding energy $E_{v}^{\textrm{b}}$ is obtained from
the solution of the Wannier equation 
\begin{equation}
\frac{\hbar^{2}k^{2}}{2m_{v}}\Psi_{v}(\mathbf{k})-\sum_{\mathbf{q}}W_{\mathbf{q}}\Psi_{v}(\mathbf{k}+\mathbf{q})=E_{v}^{\textrm{0}}\Psi_{v}(\mathbf{k})\quad,\label{Wan}
\end{equation}
where $m_{v}$ is the reduced exciton mass in valley $v$, i.e. $1/m_{v}=1/m_{v}^{e}+1/m_{v}^{h}$
with $m^{e,h}$ denoting the effective mass of electrons and holes
in the excitonic valley $v$. Here, $m_{v}$ and $E_{v}^{0}$ crucially
depend on strain. We take the starting parameters for the unstrained
single-particle dispersion from Ref. \cite{Kormanyos15} and use the
strain-dependent variation of single-particle parameters provided
in Ref. \cite{Khatibi18}. Furthermore, $\Psi_{v}(\mathbf{k})$ appearing
in Eq. (\ref{Wan}) is the excitonic wave function in momentum space,
while $W_{\mathbf{q}}$ is the Coulomb interaction, for which we have
used a modified form of the potential for charges in a thin film of
thickness $d$ surrounded by a dielectric environment \cite{Brem19b}.
Taking into account anisotropic dielectric tensors and solving the
Poisson equation with the boundary conditions described above yields
$W_{q}=V_{q}/\epsilon_{scr}(q)$ with the bare 2D-Fourier transformed
Coulomb potential $V_{q}$ and a non-local screening $\epsilon_{scr}(q)=\kappa_{1}\tanh\left[\frac{1}{2}\left[\alpha_{1}dq-\ln\left(\frac{\kappa_{1}-\kappa_{2}}{\kappa_{1}+\kappa_{2}}\right)\right]\right]$,
where $\kappa_{i}=\sqrt{\epsilon_{i}^{\parallel}\epsilon_{i}^{\bot}}$
and $\alpha_{i}=\sqrt{\epsilon_{i}^{\parallel}/\epsilon_{i}^{\bot}}$
account for the parallel and perpendicular component of the dielectric
tensor $\epsilon_{i}$ of monolayer (i = 1) and environment (i = 2),
whose parameters have been taken from Refs. \cite{Laturia18} and
\cite{Geick66}, respectively.
In this work, we consider the intrinsic undoped regime, where the effect of trions \cite{Gao16,Kato16,Cadiz17,Titze18} is negligible. Moreover,  we focus on intrinsic excitonic properties, hence we disregard defects/disorders, which are expected to lead to 
a quantitative slow-down of the overall  propagation without affecting qualitatively the strain dependence. Finally, we also neglect the impact of
spin-dark states \cite{Molas17,Cadiz17}, which are known to appear in  PL spectra under certain conditions, but since they have  similar properties as KK$^\prime$ excitons (in terms of dispersion relation), they are expected to  only lead  to quantitative changes.

The scattering-induced dynamics for the Wigner function can be written
as $\left.\dot{N}_{\mathbf{Q}}^{v}(\mathbf{r},t)\right|_{sc}=\Gamma_{\mathbf{Q}}^{\text{in},v}(\mathbf{r},t)\!-\!\Gamma_{\mathbf{Q}}^{\text{out},v}N_{\mathbf{Q}}(\mathbf{r},t)$,
where $\Gamma_{\mathbf{Q}}^{\text{in},v}(\mathbf{r},t)=\sum_{\mathbf{Q}',v'}\Gamma_{\mathbf{Q}\mathbf{Q}^{\prime}}^{vv^{\prime}}N_{\mathbf{Q}'}^{v'}(\mathbf{r},t)$
and $\Gamma_{\mathbf{Q}}^{\text{out},v}=\sum_{\mathbf{Q}',v'}\Gamma_{\mathbf{Q}'\mathbf{Q}}^{v'v}$.
The scattering rates $\Gamma_{\mathbf{Q}\mathbf{Q}^{\prime}}^{vv^{\prime}}$
of the exciton-phonon coupling  read in second-order Born-Markov approximation \cite{Brem18,kadi14}
\[
\Gamma_{\mathbf{Q}\mathbf{Q}^{\prime}}^{vv^{\prime}}=\frac{2\pi}{\hbar}\sum_{\xi\pm}\left|G_{\xi\vert\mathbf{Q}^{\prime}-\mathbf{Q}+\mathbf{Q}_{v^{\prime}}^{\prime}-\mathbf{Q}_{v}\vert}^{vv^{\prime}}\right|^{2}\left(n_{\xi,\mathbf{Q}^{\prime}-\mathbf{Q}+\mathbf{Q}_{v^{\prime}}^{\prime}-\mathbf{Q}_{v}}+\frac{1}{2}\pm\frac{1}{2}\right)\delta\left(\epsilon_{\alpha}\pm\hbar\omega_{\xi,\mathbf{Q}^{\prime}-\mathbf{Q}+\mathbf{Q}_{v^{\prime}}^{\prime}-\mathbf{Q}_{v}}-\epsilon_{\alpha^{\prime}}\right)\quad,
\]
where $\epsilon_{\alpha^{(\prime)}}\equiv E_{v^{(\prime)}}+\hbar^2|\mathbf{Q}^{(\prime)}|^2/2M_{v^{(\prime)}}$ and $\mathbf{Q}_{v}$ addresses the position in the Brillouin zone
of the bottom of valley $v$, $\pm$ refers to emission/absorption
of phonons, $\xi$ labels the phonon modes, $\hbar\omega_{\xi\mathbf{q}}$
provides the energy of phonon $\xi$ with momentum $\mathbf{q}$ and
$n_{\xi,\mathbf{q}}$ describes the equilibrium phonon distribution. We take into account  longitudinal and transverse acoustic (LA, TA) and optical (LO, TO) modes as well as  
the out-of-plane A$_1$ optical mode, which provide the most efficient scattering channels 
\cite{Jin14}.
The exciton-phonon scattering coefficients $G$ can be written as
\cite{Brem19b} $\left|G_{\xi\mathbf{q}}^{vv^{\prime}}\right|^{2}=\sum_{p=e,h}\left|g_{p\xi\mathbf{q}}^{vv^{\prime}}\mathcal{F}_{p}^{vv^{\prime}}(\mathbf{q})\right|^{2}$,
where $p$ distinguishes between electrons/holes, $\mathcal{F}$ is
the form factor $\mathcal{F}_{p}^{vv^{\prime}}(\mathbf{q})=\sum_{\mathbf{k}}\Psi_{v^{\prime}}^{*}(\mathbf{k})\Psi_{v}(\mathbf{k}+\alpha_{v}^{p}\mathbf{q})$,
with $\alpha_{v}^{p}=(m_{v}^{e}+m_{v}^{h})^{-1}\left[\pm m_{v}^{h/e}\right]$
for $p=e,h$, respectively, while $g$ are the carrier-phonon scattering
coefficients which are approximated with the generic form of a deformation
potential 
$
g_{p\xi\mathbf{q}}^{vv^{\prime}}=\sqrt{\frac{\hbar}{2\rho A\omega_{\xi\mathbf{q}}}}\mathcal{D}_{p\xi\mathbf{q}}.
$
Here, $\rho$ denotes the surface mass density of the monolayer (whose
small variations with strain have been omitted) and $A$ the area
of the system.  For the coupling constant $\mathcal{D}_{p\xi\mathbf{q}}$
we adopt the approximations deduced from DFPT calculations in Ref.
\cite{Jin14}, where long range acoustic phonons couple linear in
momentum $\mathcal{D}_{\xi\mathbf{q}}\equiv\mathcal{D}_{\xi}^{(1)}q$,
while optical phonons and short range acoustic modes couple with a
constant strength $\mathcal{D}_{\xi\mathbf{q}}\equiv\mathcal{D}_{\xi}^{(0)}$
in the vicinity of high symmetry points. The phonon energies as well
as deformation constants $\mathcal{D}^{(0)}$ and $\mathcal{D}^{(1)}$
for all possible intra and intervalley scattering channels are listed
in Ref. \cite{Jin14}, including longitudinal and transverse acoustic
(LA, TA) and optical (LO, TO) modes as well as the out-of-plane A$_{1}$
optical mode.
Finally, for describing the radiative decay \cite{Brem19b,Selig18}
we use $\gamma=4.5$ meV in accordance to a recombination-induced
broadening of the linewidth of 4.5 meV in FWHM \cite{Khatibi18}. We find, however,  that the radiative recombination rate has a negligible
role on the qualitative aspects of the investigated effective diffusion
in h-BN encapsulated WS$_{2}$ monolayers \cite{Rosati20}. For the
same reason we have also neglected strain-induced changes of $\gamma$,
which are typically weak \cite{Khatibi18,Aslan18}.

\section{Exciton diffusion in WS$_{2}$ at lower temperatures}

\begin{figure}[t!]
\centering \includegraphics[width=10cm]{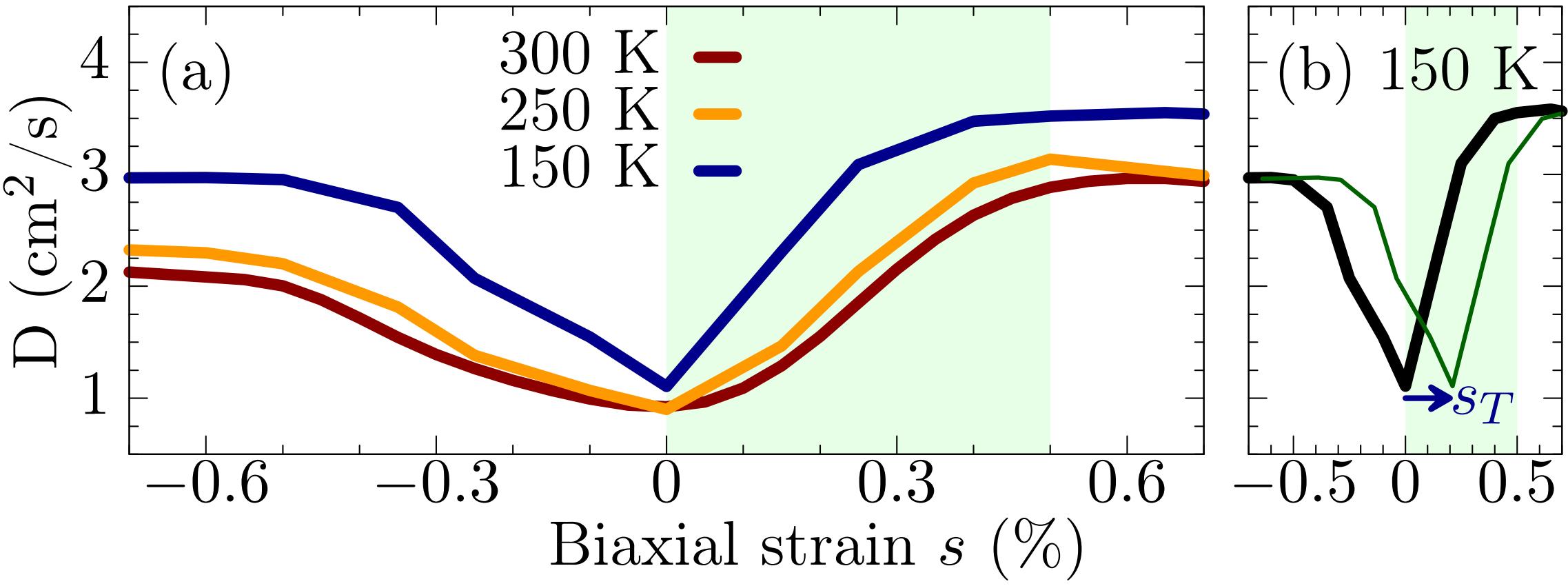} \caption{\textbf{Impact of temperature.} (a) Strain-dependent stationary diffusion
coefficients at different temperatures and (b) at 150 K assuming an
additional background strain $s_{T}$ due to the lattice contraction
at smaller temperatures. The green-shaded area indicates the experimentally
accessible strain region at room temperature.}
\label{fig:Supp1} 
\end{figure}

We study the impact of temperature on the predicted dip
in the exciton diffusion for unstrained WS$_{2}$, cf. Fig. 3(a) in
the main text. A decrease in temperature is accompanied by a spectral
shift of excitonic resonances, often described via the Varshni equation
\cite{Varshni67}. The shift originates from an interplay
of temperature-induced changes in the electron-phonon interaction (polaron shift)
and a the lattice contraction at smaller temperature. We denote the
latter effect as temperature-induced strain $s_{T}$. To which extent
the shift is due to one or the other contribution is still under debate
in literature, see e.g. \cite{Odonnell91} and references therein.
This goes beyond the scope of the present work. For the moment, we neglect this effect, i.e. $s_{T}=0$, and we show the exciton
diffusion as a function of temperature in Fig. \ref{fig:Supp1}(a).
We find that the predicted dip in the diffusion becomes narrower for
decreasing temperatures. This can be easily explained by studying
Figs. 4(a)-(c) in the main manuscript: While the position of the thin
red line is assumed independent of temperature, the width of the exciton
occupation in momentum space decreases with temperature. As a result,
at smaller temperatures already at $s=$0.25 $\%$ the intervalley scattering
via absorption of TA modes is suppressed due to a narrower distribution
in momentum, and at this strain value the corresponding dip already
saturates, as shown in Fig. \ref{fig:Supp1}(a).

Interestingly, the minimum of the dip at $s\approx0\%$ seems independent
of temperature. To understand this we study the diffusion coefficient
from the Fick's law by considering only one phonon mode and assuming
$\tau_{Q}\equiv\tau_{\text{ph}}$. The resulting diffusion $D_{\text{ph}}\equiv\tau_{\text{ph}}k_{B}T/M$
behaves with temperature as $D_{\text{ph}}\propto\text{Exp}\left({\frac{\hbar\omega_{\text{ph}}}{k_{B}T}}\right)T\equiv f_{\text{ph}}(T)$.
In the range of temperatures between 150 and 300 K $f_{\text{TA}}$
varies weakly. This is not the case anymore when considering more
energetic phonon modes. Here, $f_{\text{TO}}$ would increase at smaller
temperatures, as expected from the naive consideration that weaker
scattering leads to a faster diffusion. Since acoustic phonons are crucial
for the generation of the dip (as discussed in the main manuscript),
we have an almost temperature-independent dip minimum in the considered
temperature range.

Finally, in Fig. \ref{fig:Supp1}(b) we consider a non-zero background strain $s_{T}$ such that it gives rise to the half of the observed temperature-induced shift
\cite{Nagler18}. This results
in a shift of the dip in the exciton diffusion to higher strain values. This 
means that the non-monotonicity appears in the strain range that is experimentally
accessible at room temperature. Although with state-of-the-art techniques
it is complicated to induce strain at smaller temperatures, Fig. \ref{fig:Supp1}(b)
predicts that the interplay between temperature-induced lattice shrinking
and additional strain could in principle lead to observations of a
pronounced dip in the exciton diffusion also at experimentally accessible tensile strain values.

\section{Exciton diffusion in other TMDs}

\begin{figure}[t!]
\centering \includegraphics[width=\columnwidth]{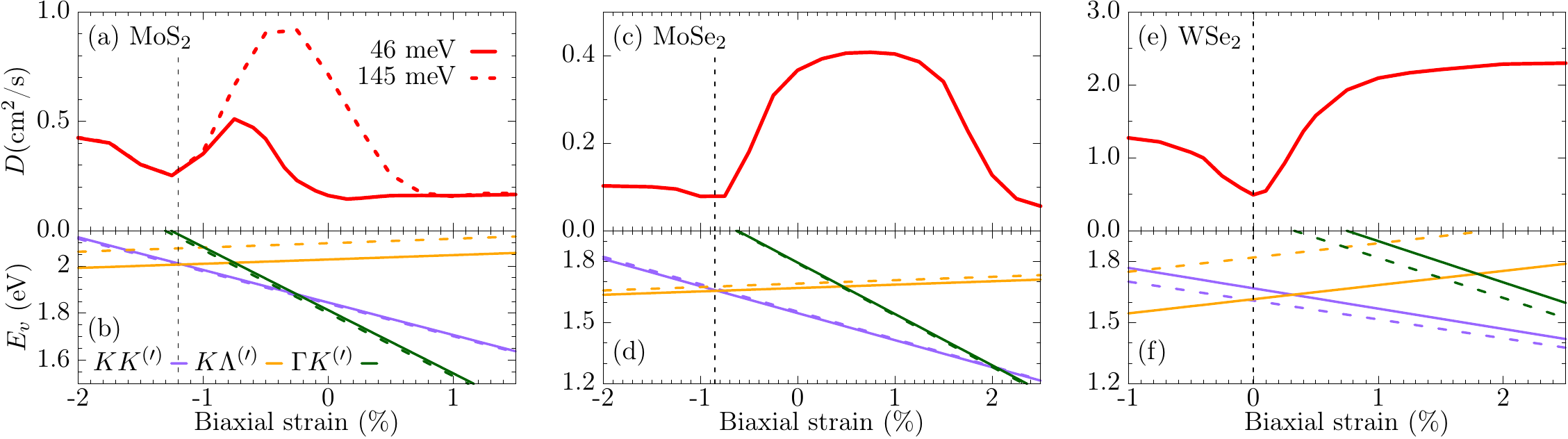} \caption{\textbf{Strain-dependent stationary diffusion in MoS$_{2}$, MoSe$_{2}$, and WSe$_{2}$.} The
same plot as in Fig. 3 in the main manuscript. In panel (a), two different values for the separation between the higher-lying valence-band maximum in the K  valley and the lower-lying one in the  $\Gamma$ valley have been considered, i.e. 46 meV \cite{Kormanyos15} (solid line) and 145 meV \cite{Khatibi18} (dashed line). In panel (b) the smaller value has been considered. }
\label{fig:Supp2} 
\end{figure}

We now briefly discuss the impact of strain on exciton diffusion in other TMD materials than discussed in the main manuscript.
In MoS$_{2}$, the separation between the valence-band maxima located at the $\Gamma$ and the K valley is still being debated in literature, cf. \cite{Kormanyos15,Khatibi18,Uchiyama19,Deilmann19}. Assuming a small separation of 46 meV \cite{Kormanyos15}, we find that already at $s\approx0$ the diffusion
is saturated, cf. the solid line in Fig. \ref{fig:Supp2}(a). This can be led back to the vicinity of the
$\Gamma$ and the K valley in the single-particle dispersion for the
unstrained case, reflecting the close position of the excitonic energies shown in Fig. \ref{fig:Supp2}(b). As a consequence,
for tensile strain almost all  excitons are located in the
$\Gamma$K$^{(\prime)}$ state. This means
that the exciton diffusion is almost unchanged for
experimentally accessible strain values between 0 and 0.5$\%$.  The behaviour is more involved for compressive strain, where we observe the maximum in the region where KK$^\prime$ excitons
are clearly energetically lowest states. Exciton diffusion further decreases for higher values of compressive strain, i.e. moving toward
the regime, where K$\Lambda^{(\prime)}$ excitons are dominant. 

The situation for tensile strain $\lesssim 1$\% values changes, if a larger $\Gamma$-K separation  is assumed. A value of 145 meV \cite{Khatibi18} leads to a decrease of the diffusion coefficient upon increase of strain, cf. the dashed red line in Fig. \ref{fig:Supp2}(a). This occurs in view of the larger separation between $E_{\Gamma\text{K}^{(\prime)}}$ and
$E_{\text{KK}^{(\prime)}}$ with the former being now able to affect the overall diffusion only if larger values of strain are applied. The increased separation also affects the maximum value of the diffusion as obtained for small
compressive strain, which is now approximately two times larger than in the case of the smaller separation, again reflecting a smaller contribution stemming from the dark $\Gamma$K and $\Gamma$K$^\prime$ excitons.

For MoSe$_2$, exciton diffusion also changes only weakly
for small tensile strain (Fig. \ref{fig:Supp2}(c)), however the microscopic origin is very different w.r.t. the solid line in Fig. \ref{fig:Supp2}(a), as can be seen in the excitonic energies shown in Fig. \ref{fig:Supp2}(d). At around 0\%,
 KK$^\prime$ excitons are energetically lowest states and dominate the diffusion. Only at larger tensile strain values ($s \approx 2 \%$) $\Gamma\text{K}^{(\prime)}$ excitons become dominant giving rise to a decrease of the diffusion coefficient -  similar to the behaviour at slightly smaller tensile strain discussed for WS$_2$ in the main manuscript. Note that in the strain regimes where the overall diffusion is dominated by KK$^{(\prime)}$ the absolute value of the diffusion coefficient is much smaller than in the analogous situation in WS$_2$ [see maximum at 0.6\% of Fig. 3(a) of the main manuscript], 
reflecting in particular a much stronger effectiveness of the exciton-phonon scattering with intravalley modes in MoSe$_2$ w.r.t. WS$_2$ \cite{Khatibi18}.
 
Finally, we find that WSe$_2$  has a similar strain behaviour as WS$_2$ for compressive strain  and small tensile strain values, cf. Fig. \ref{fig:Supp2}(e). However, the decrease of the diffusion coefficient at larger  tensile strain
does not take place due to different energy positions  of $E_{\Gamma\text{K}^{(\prime)}}$ and $E_{\text{KK}^{(\prime)}}$, Fig. \ref{fig:Supp2}(f).

\section{Experimental determination of the diffusion coefficients of strained WS$_{2}$ monolayers}

\subsection{Strained WS$_{2}$ monolayers}

\begin{figure}[t!]
\centering \includegraphics[width=14cm]{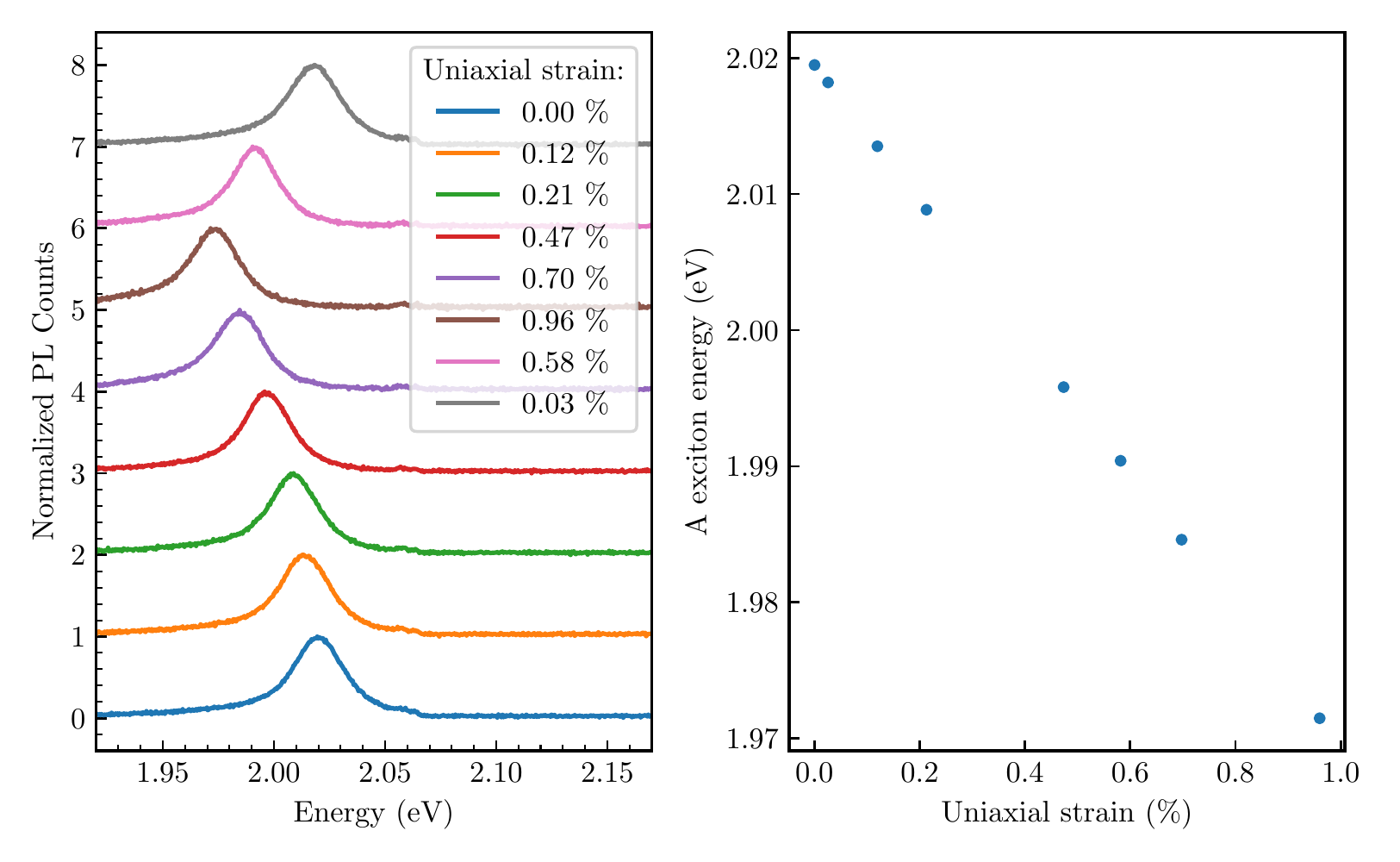} \caption{\textbf{Photoluminescence of the WS$_{2}$ monolayer under uniaxial
tensile strain.} (a) Measured photoluminescence spectra for different
uniaxial strain values. (b) Strain-dependent A exciton energies extracted from (a).}
\label{fig:Supp3} 
\end{figure}

To investigate the exciton diffusion in strained WS$_{2}$, a monolayer
is mechanically exfoliated onto a Polymethyl methacrylate (PMMA) substrate
(R\"ohm GmbH Plexiglas 0F058) of 0.5~mm thickness. By bending the substrate,
the monolayer on top is uniaxially strained by up to 0.96~\% \cite{Schmidt16}.
Excellent strain transfer is verified by the observed red-shift of
the photoluminescence (Fig. \ref{fig:Supp3}) with a gauge factor
of 50~meV/\%. To directly compare the experimental and theoretical
data, the experimental uniaxial strain values have to be related to
the biaxial values from the theory. Both uniaxial and biaxial tensile
strain cause similar energetic shifts of the excitonic transitions,
and only their gauge factors differ. For exciton diffusion, the relative energetic
shift of  excitonic transitions is the relevant parameter. Therefore,
those experimental uniaxial strain values correspond to the theoretical
biaxial ones, which yield the same exciton shift. To this end, the
experimental strain values are multiplied by the ratio of the measured
uniaxial gauge factor and the calculated biaxial gauge factor [cf. Fig. \ref{fig:Supp3}(b)]
\begin{align}
\frac{\alpha_{\text{uniaxial}}}{\alpha_{\text{biaxial}}}=\frac{50~\text{meV/\%}}{113~\text{meV/\%}}=0.44.
\end{align}
In this way, the experimental and theoretical gauge factors match,
i.e. the experiment spans a biaxial strain range from 0~\% to 0.42~\%.

\subsection{Exciton diffusion measurement and analysis}

\begin{figure}[t!]
\centering \includegraphics[width=08cm]{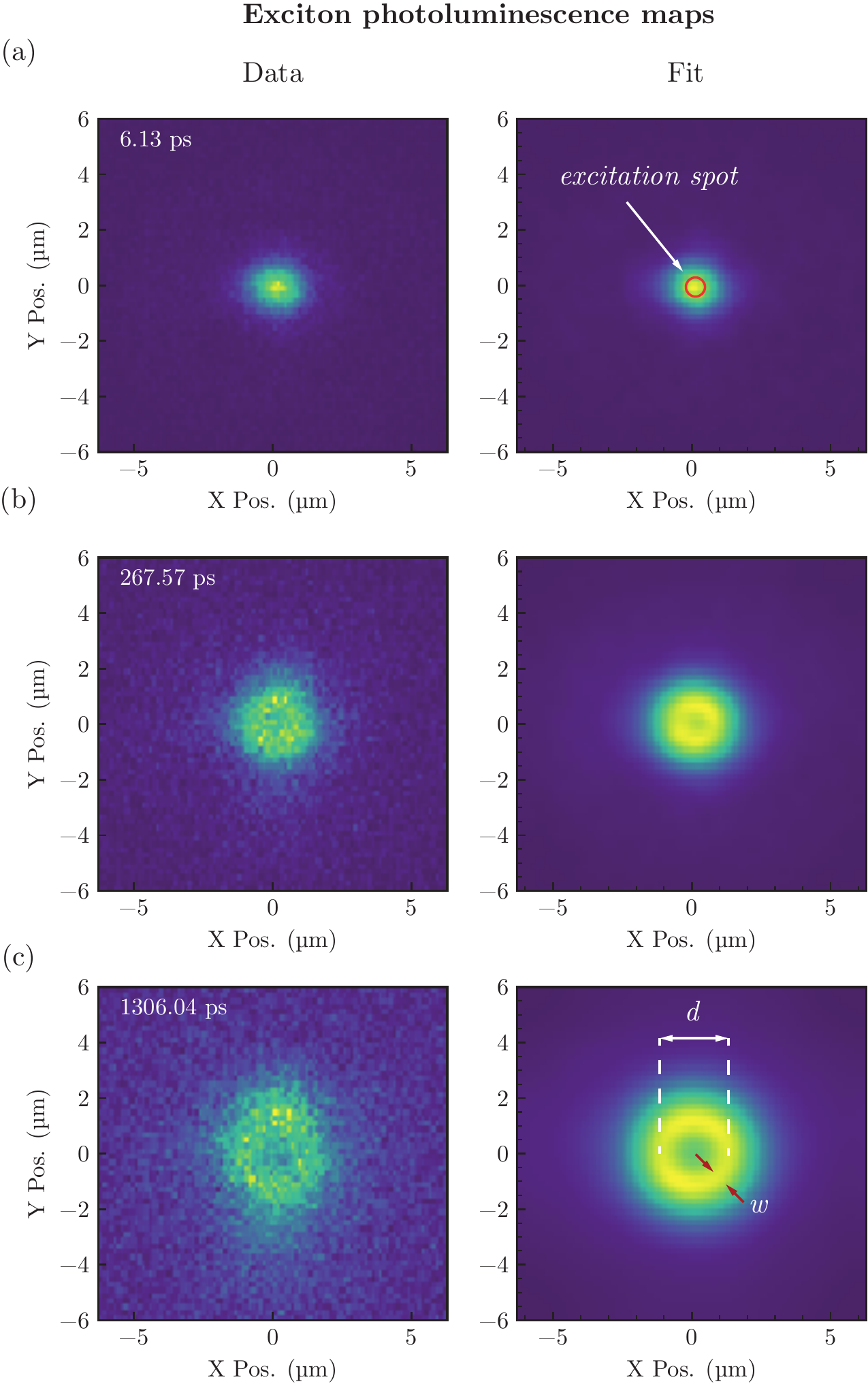} \caption{\textbf{Exciton photoluminescence maps for different delay times.}
(a)-(c) Exemplary PL maps of a strained WS$_{2}$ monolayer for different
delay times after femtosecond excitation. The left column depicts
the experimental data and the right column the fit with a donut-like
function. A strain direction-dependent effect on the diffusion is not observable.}
\label{fig:Supp4} 
\end{figure}

To measure exciton diffusion, the monolayer is optically excited
using 250~fs pulses at 2.10~eV from a 40~MHz femtosecond fiber
laser system \cite{Schmidt16b}. The monolayer PL is measured in a
transmission microscope using a 50x ($\mathrm{NA}=0.55$) objective
lens for excitation, resulting in a focus size of 700~nm at the sample.
For detection, we use a 100x ($\mathrm{NA}=0.9$) objective lens.
A streak camera (Hamamatsu C10910-05) detects the photoluminescence
with a temporal resolution in the picosecond range. The detection spot of the PL can be scanned across the sample
by moving the detection objective lens. 
The residual pump light is blocked using a long-pass filter at 2.07 eV.
From the measurement we reconstruct an image sequence of
the spatial distribution of the monolayer photoluminescence with a
temporal resolution of $\approx$ 5~ps and a temporal range of up
to 2~ns (Fig. \ref{fig:Supp4}).

Kulig et al. \cite{Kulig18} have demonstrated that exciton diffusion
in monolayer WS$_{2}$ depends on the excitation density, except for
very low powers in the nanowatt regime. For higher powers, so-called
``halos'' are observable in the exciton photoluminescence. To be
able to observe the impact of strain on  diffusion coefficients
predicted by the theory, exciton diffusion has to be measured in the
regime, where it is independent of the excitation density.

We use two different approaches to measure exciton diffusion. In the
first, we are in the excitation regime, where  exciton diffusion shows
the formation of halos (excitation power 2~\textmu W). In this case, excitons
are quickly pushed away from the irradiated into a pristine area of
the substrate, where they can diffuse freely.

The experimental data is fitted using a donut-like function (Fig. \ref{fig:Supp4}):
\begin{align}
I_{\text{PL}}(x,y)=A_{0}\cdot\left(\text{e}^{-\left((x-x_{0})^{2}+(y-y_{0})^{2}-d/2\right)^{2}/2\sigma^{2}}+\text{e}^{-\left((x-x_{0})^{2}+(y-y_{0})^{2}+d/2\right)^{2}/2\sigma^{2}}\right)
\end{align}
Here, 
$I_{\mathrm{PL}}$ is the intensity of the photoluminescence at the
specific coordinate, $x_{0}$ and $y_{0}$ denote the center of the donut,
$A_{0}$ the amplitude, $d$ the diameter and $\sigma$ the width
of the ring. In this excitation regime, exciton transport consists
of an excitation density-dependent part, creating the halo and the
density-independent exciton diffusion. The first part dominates the
temporal evolution of the parameter $d$, while the latter one can
be extracted from the increasing width of the halo ring $w$, especially
at long delay times, when the exciton density has already decreased
by 1 to 2 orders of magnitude.
The time evolution of $w^2=2\sigma^2$ is analyzed for delay times \textgreater{}
700~ps and the diffusion coefficient $D$ can be calculated as follows:
\begin{align}
D & =\frac{w^{2}-w_{700\mathrm{ps}}^{2}}{4(t-t_{700\mathrm{ps}})} \quad .
\end{align}

\begin{figure}[t!]
\begin{centering}
\includegraphics[width=12cm]{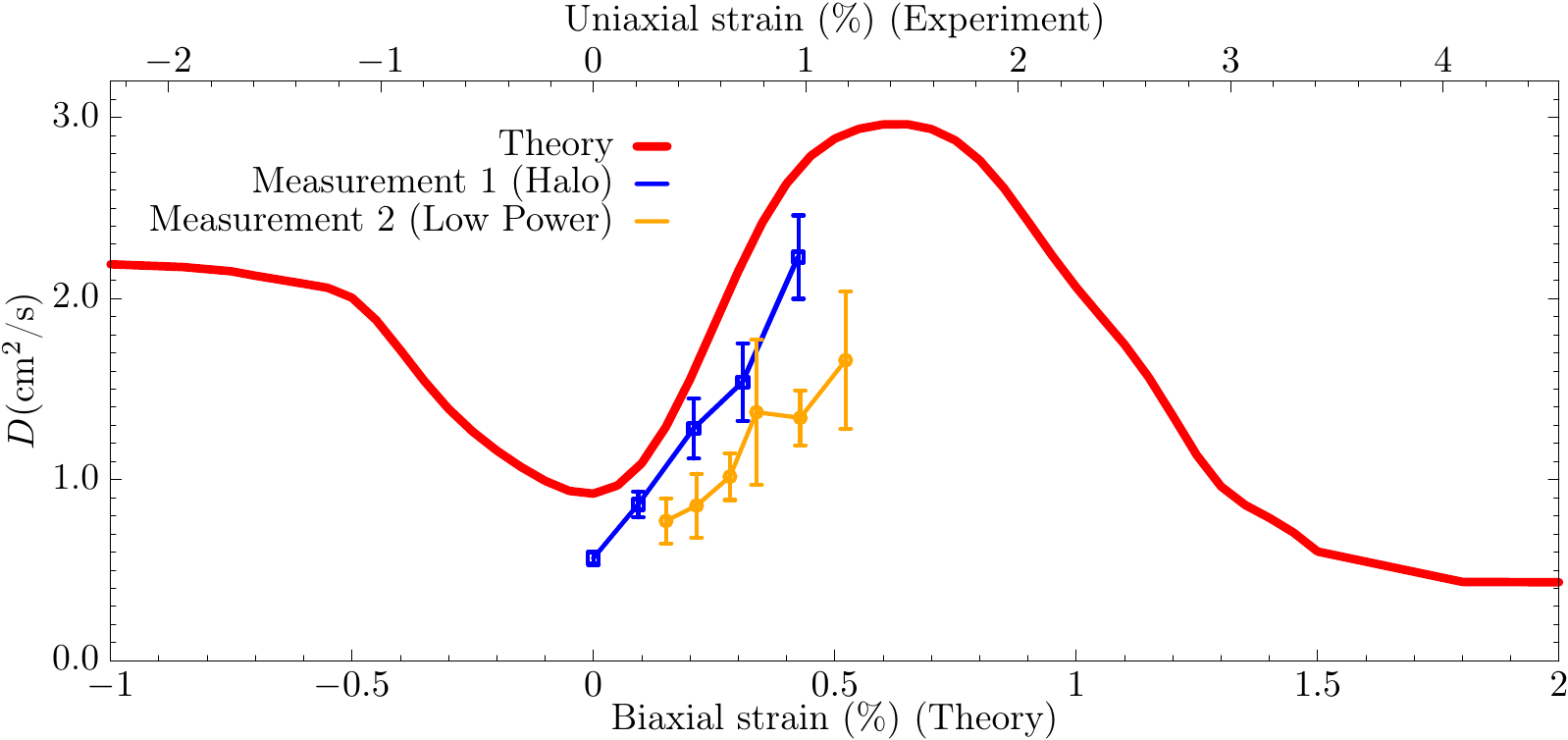}
\par\end{centering}
\caption{Comparison of the experimental diffusion coefficients (measured with
the two approaches described in the text) and the theorical prediction.}

\label{fig:Supp5}
\end{figure}

In the second approach, we use the low excitation power regime, where
 exciton diffusion is independent of the excitation density (excitation
power 5~nW) and no halo formation is visible. To decrease the measurement time we measure the diffusion only in one dimension
instead of the entire PL image, i.e. we measure a horizontal cut through
the center of the excitation spot, which has a Gaussian shape (no
halo). Therefore, the experimental data can be modeled by a Gaussian
function with the width $w$ and the diffusion coefficient
$D$ can be calculated by
\begin{align}
D & =\frac{w^{2}-w_{0\mathrm{ps}}^{2}}{4t} \quad .
\end{align}
Figure \ref{fig:Supp5} presents the results of the two measurement
approaches and compares them to the theoretical prediction. Both approaches yield
the same qualitative trend and show a very good agreement with the theory.

